# Quantum Encryption in Phase Space using Displacement Operator for QPSK Data Modulation

Randy Kuang
Quantropi Inc.
Ottawa, Canada
randy.kuang@quantropi.com
ORCID: 000-0002-5567-2192

Adrian Chan
Quantropi Inc.
Ottawa, Canada
adrian.chan@quantropi.com

*Abstract*— **In 2020, Kuang and Bettenburg proposed Quantum Public Key Distribution (QPKE) which utilized the randomized phase shift gate. Since then, it has been implemented both theoretically through simulations and experimentally over existing fiber optical networks. QPKE can be compared to an RSA-type scheme but in the optical analogue domain. Later on, it was renamed Quantum Encryption in Phase Space (QEPS) to emphasize the encryption of coherent states in phase space. However, the phase shift gate used in QEPS is limited to data modulation schemes with phase shift keying such as quadrature phase shift keying (QPSK) as it may leak data information in amplitude if applied to quadrature amplitude modulation (QAM) schemes. Recently, Kuang and Chan proposed a new version of QEPS known as Quantum Encryption in Phase Space with the displacement gate or QEPS-d, which overcomes the limitation of QEPS with the phase shift gate. This was achieved by introducing a reduced displacement operator that ignores the global phase factor, making the reduced displacement operators commutable, thus aiding the implementation at both transmission and reception. Furthermore, any arbitrary displacement operator can be decoupled into a standard QAM modulation with a phase shift modulation, making encryption and decryption easier. This paper demonstrates the simulation of QEPS-d encryption for QPSK data modulation to illustrate how QEPS-d functions.**

## I. Introduction

In 1994, Shor proposed an algorithm using quantum bits (qubits) for integer factorization [1], which highlighted the vulnerability of classical public key algorithms such as RSA, Diffie-Hellman, and elliptic Diffie-Hellman based on factorization and discrete logarithm problems, respectively, to fault-tolerant quantum computers. However, breaking RSA-2048 requires a fault-tolerant quantum computer with over 4000 logic qubits or 4 million physical qubits. The latest IBM quantum computer, Osprey, has 433 physical qubits [2]. According to the IBM roadmap, the next quantum computer, Condor, with 1121 qubits, will be released in 2023, and the number of qubits is expected to exceed 100,000 by 2026.

Recently, Yan et al. proposed a new algorithm called Sublinear-resource Quantum Integer Factorization (SQIF) [3]. SQIF combines classical lattice reduction with quantum optimization and can operate on a noisy quantum computer, reducing the quantum resource from 4 million physical qubits to less than 400 physical qubits. The researchers demonstrated this by factorizing a 48-bit integer with as little as a 10-qubit quantum processor.

Starting in late 2017, the National Institute of Standards and Technology (NIST) began the standardization process for key encapsulation mechanisms (KEM) and digital signature algorithms. After completing three rounds of review, NIST announced the final standardized algorithms in 2021 [4]. For KEM, the lattice-based Kyber [6] emerged as the winner, while for digital signatures, NIST standardized the lattice-based Dilithium [7] and Falcon [8], as well as the hash-based SPHINCS+ [9]. Moving forward, NIST will continue its standardization process for KEM in round 4. Additionally, in early 2023, NIST plans to reopen the standardization process for digital signature submissions.

In 2022, some major cryptanalyses revealed vulnerabilities in several of the NIST finalists. For instance, Beullens was able to break the Rainbow signature using only a laptop over a single weekend [10]. Robert broke SIDH [11], while Castryck and Decru developed a more efficient method to break SIDH level I using a single-core computer, achieving success within an hour [12]. Additionally, Wenger et al. reported their successful recovery of a lattice-based PQC using machine learning. By training a transformer with 300,000 samples [13], they were able to achieve complete secret recovery for lattice dimensions up to a mid-size range.

Kaung's team recently proposed Multivariate Polynomial Public Key (MPPK) for post-quantum cryptography (PQC) key encapsulation mechanism (KEM) and digital signature, leveraging the NP-complete problem of the Modular Diophantine Equation Problem [14, 15, 16, 17]. MPPK offers relatively small public key, cipher, and signature sizes, comparable to classical public key schemes. It also outperforms NIST finalists in key generation, encryption, decryption, signing, and verification. As such, MPPK could be a good alternative to NIST finalists for generic use cases. MPPK's digital signature scheme is expected to participate in the NIST reopening submission for digital signature.

Meanwhile, Quantum Key Distribution (QKD) has been developed over the last three decades since it was proposed in 1984. In 2000, Shor and Preskill proved that QKD offers information-theoretical security [18]. It has become commercially ready for distances of around 100km. To break the distance barrier, Lucamarini et al. proposed Twin-Field QKD (TF-QKD) in 2018 [19]. Since then, TF-QKD has been widely explored, and in 2022, Wang et al. reported the longest distance of 830km [20]. Generally, QKD offers a key rate at the kbps level, and TF-QKD achieved a key rate of 0.014 bps at 830km, taking more than 5 hours to establish a 256-bit AES key.

Considering the pre-shared secret for QKD authentication, Kuang and Bettenburg in 2020 proposed a new mechanism using Quantum Permutation Pad or QPP to digitally distribute quantum random [21]. The pre-shared secret is not only used for authentication but also used to map to a QPP pad for encoding at the sender and decoding at the receiver. QPP is implemented into matrices operating on data column vector or Dirac ket. Permutation matrix is unitary and reversable, so the decoding side uses the reversed QPP. Kuang and Barbeau proposed a universal quantum safe cryptography using QPP in 2022 [22]. QPP has been developed as a platform for digital QKD and benchmarked by Deutsche Telekom in 2022 [23]. Leveraging the quantum gate property of QPP, quantum encryption with QPP implemented inside quantum computers was reported by Kuang and Perepechaenko in 2022 [24], Perepechaenko and Kuang in 2022 [25, 26].

To eliminate the pre-shared key in quantum key distribution in coherent optical domain, Kuang and Bettenburg in 2020 proposed Quantum Public Key Envelope or QPKE using randomized phase shift gate in a round-trip scheme [27], leveraging the self-shared random secret to drive the phase shift encoding without the specific requirement of the pre-shared secret. QPKE was designed to operate in the existing coherent optical networks with the same coherent detection module. It has been simulated and experimentally implemented through the collaborations with McGill University [28, 29, 30, 31]. QPKE mimics the RSA-type public key scheme in coherent optical domain. The experiment implementation with off-shelf optical modules demonstrated the speed at 200 gbps for a distance 80km between two communication peers. To mimicking its implementation in a symmetric fashion with a pre-shared secret, QPKE was renamed as Quantum Encryption in Phase Space or QEPS with the randomized phase shift gate, reflecting to its possible implementation in photonic quantum computer with phase shift gate. There is one limitation of QEPS with phase shift gate, or only applicable for data modulation schemes with phase shift keying such as QPSK or M-PSK. Once the data modulation is quadrature amplitude modulation or QAM, the amplitude bits would be leaked out.

To overcome this limitation, Kuang and Chan recently proposed to use coherent displacement operator $\widehat{D}(\alpha)$ where $\alpha$ denotes a coherent state [32]. This paper will report its simulation results with QPSK data modulation. Section 2 will briefly summarize the QEPS with the displacement operator and section 3 will present the simulation result and the conclusion is at the end.

In 2020, Kuang and Bettenburg introduced Quantum Public Key Envelope (QPKE) to remove the need for pre-shared keys in quantum key distribution within the coherent optical domain [27]. They employed a randomized phase shift gate in a round-trip scheme to utilize a self-shared random secret for phase shift encoding, eliminating the need for a pre-shared secret. QPKE is compatible with existing coherent optical networks and can be simulated and implemented in collaboration with McGill University [28, 29, 30, 31]. QPKE resembles an RSA-type public key scheme in the coherent optical domain and can achieve a speed of 200 gbps between two communication peers over a distance of 80 km using off-the-shelf optical modules. To implement QPKE in a symmetric fashion with a pre-shared secret, it was renamed Quantum Encryption in Phase Space (QEPS) with a randomized phase shift gate. However, the use of the phase shift keying data modulation scheme limits the applicability of QEPS to only QPSK or M-PSK data modulation schemes, as the amplitude bits would be leaked out if the data modulation were quadrature amplitude modulation or QAM. To address this issue, Kuang and Chan proposed using a coherent displacement operator $\widehat{D}(\alpha)$, where α represents a coherent state [32].

On one hand, there are proposed other optical encryption methods that require significant modifications to the optical transceivers, such as optical encryption based on tightly focused perfect optical vortex beams by Yang et al. in 2022 [33], a new optical coherence encryption with structured random light introduced by Peng et al. in 2021 [34], which encodes information into the second-order spatial coherence distribution of a structured random light beam, and Jaramillo-Osorio et al.'s 2022 optical encryption method using phase modulation generated by thermal lens effect [35]. On the other hand, the QEPS approach is designed to make use of off-the-shelf optical modules or, if necessary, software and firmware updates in commercial optical transceivers.

The paper summarizes QEPS with the displacement operator in Section 2 and presents the simulation results with QPSK data modulation in Section 3, with the conclusion provided at the end.

## II. QEPS WITH DISPLACEMENT OPERATOR

### A. Coherent State and Displacement Operator

A coherent state is the specific quantum state of quantum harmonic oscillator denoted by a Dirac ket $|\beta\rangle$ where $\beta$ is a complex variable in the phase space. $\beta$ can be expressed either in terms of in-phase and quadrature as $\beta = \beta_i + j\beta_q$ or amplitude and phase $\beta = |\beta|e^{j\varphi}\rangle$. Then a coherent state can be written as

$$|\beta\rangle = |\beta_i + j\beta_q\rangle = ||\beta|e^{j\varphi}\rangle \quad (1)$$

And the displacement operator is defined with creation and annihilation operators $\hat{a}^\dagger$ and $\hat{a}$ through following equation

$$|\alpha\rangle = e^{\alpha\hat{a}^\dagger - \alpha^*\hat{a}}|0\rangle = \widehat{D}(\alpha)|0\rangle \quad (2)$$

So

$$\widehat{D}(\alpha) = e^{\alpha\hat{a}^\dagger - \alpha^*\hat{a}} \quad (3)$$

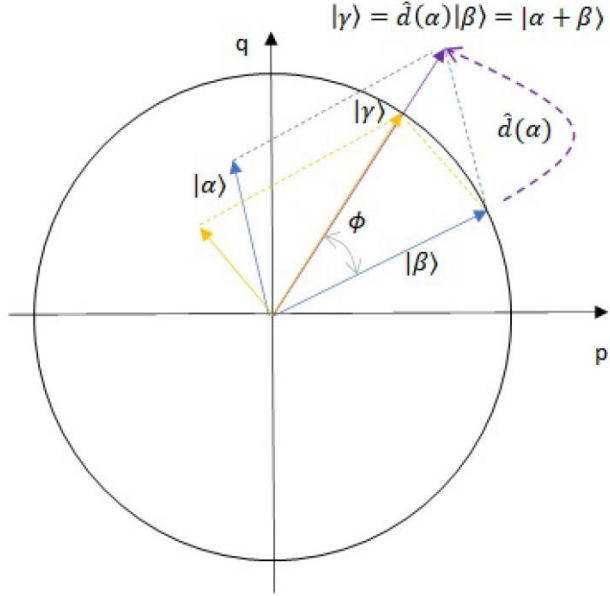

Figure 1. Illustration of QEPS-d is plotted in the phase space. A special case of QEPS with phase shift operator is also plotted for demonstration purpose of a general displacement operator $\hat{d}(\alpha)$.

which indicates the displacement operator is unitary and reversable:

$$\hat{D}^{\dagger}(\alpha) = \left(e^{\alpha \hat{a}^{\dagger} - \alpha^* \hat{a}}\right)^{\dagger} = \hat{D}(-\alpha) = \hat{D}^{-1}(\alpha) \quad (4)$$

Let's apply the displacement operator $\hat{D}(\alpha)$ to a coherent state $|\beta\rangle$

$$\hat{D}(\alpha)|\beta\rangle = \hat{D}(\alpha)\hat{D}(\beta)|0\rangle = e^{\alpha\beta^* - \alpha^*\beta}\hat{D}(\alpha + \beta)|0\rangle \quad (5)$$

And in the same way

$$\hat{D}(\beta)|\alpha\rangle = \hat{D}(\beta)\hat{D}(\alpha)|0\rangle = e^{\beta\alpha^* - \beta^*\alpha}\hat{D}(\alpha + \beta)|0\rangle \quad (6)$$

So, it is clear that $\hat{D}(\alpha)$ and $\hat{D}(\beta)$ are not commutable due to the global phase factor $e^{\alpha\beta^* - \alpha^*\beta}$ but that does not impact our physical measurements on the amplitude and phase of a coherent state. Therefore, we can ignore the global phase factor and introduce a reduced displacement operator $\hat{d}(\alpha) = e^{-\alpha\beta^* + \alpha^*\beta}\hat{D}(\alpha)$. Then the reduced displacement operator $\hat{d}(\alpha)$ and $\hat{d}(\beta)$ are commutable.

### B. QEPS with Reduced Displacement Operator

From Eq. (5), QEPS encryption with a reduced displacement operator $\hat{d}(\alpha)$ can be expressed as follows:

$$\hat{d}(\alpha)|\beta\rangle = \hat{d}(\alpha + \beta)|0\rangle = |\alpha + \beta\rangle = |\gamma\rangle \quad (7)$$

with $|\beta\rangle$ to be a plain coherent state, $\hat{d}(\alpha)$ to be an encryption operator and $|\gamma\rangle$ to be the encrypted cipher coherent state. Eq. (7) indicates that QEPS encryption with the reduced displacement operator or QEPS-d essentially performs an addition of two coherent states $|\alpha\rangle$ and $|\beta\rangle$ as shown in Fig. 1. A general displacement operator would change both the amplitude and phase of a plain coherent state. But it can also only change the phase of the plain coherent as shown in Fig. 1.

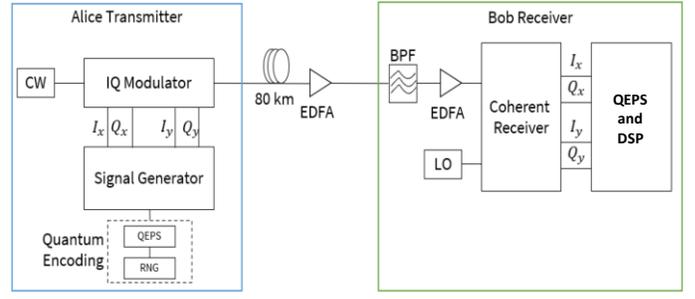

Figure 2. Simulation layout is illustrated. CW: continuous wave source, IQ Modulator: in-phase and quadrature modulator, $I_x, Q_x$ and $I_y, Q_y$: in-phase and quadrature components for IQ modulator, QEPS: coherent encryption module driven by a random number generator or RNG seeded with a pre-shared secret, EDFA: Erbium-Doped Fiber Amplifier, Coherent Receiver: coherent detection, LO: local oscillator, QEPS and DSP: digital QEPS decryption and DSP.

In this special case, the displacement operator behaves like a phase shift operator.

The encryptor $\hat{d}(\alpha)$ can be controlled by a pre-shared secret in a symmetric encryption or a self-shared secret in an asymmetric encryption as shown in QPKE [27]. In the ideal communication case, the receiver would decrypt the cipher coherent state $|\gamma\rangle$ with $\hat{d}^{-1}(\alpha) = \hat{d}(-\alpha)$ : $\hat{d}^{-1}(\alpha)|\gamma\rangle = \hat{d}(-\alpha)|\gamma\rangle = |-\alpha + \gamma\rangle = |\beta\rangle$.

In coherent optical communications, optical line path would impact a coherent state during transmission from the sender to the receiver such as dispersion, attenuation, polarization, noise, environment factors, etc. Thanks to the digital signal processing or DSP, all those impacts could be compensated and corrected in the electrical digital domain. Based on that, we only consider the encryption and decryption in the ideal transmission situation.

A displacement operator can be decomposed into two or more displacement operators as follows

$$\hat{d}(\alpha) = \hat{d}(\alpha_1)\hat{d}(\alpha_2)\ldots\hat{d}(\alpha_m)$$

And

$$\hat{d}(\alpha)|\beta\rangle = \hat{d}(\alpha_1)\hat{d}(\alpha_2)\ldots\hat{d}(\alpha_m)|\beta\rangle$$
$$= |\alpha_1 + \alpha_2 + \cdots \alpha_m + \beta\rangle$$

This decomposition feature helps us to ease the implementation of a general displacement operator with two operators: $\hat{d}(\alpha_1)$ implemented with a standard modulation such as QAM and $\hat{d}(\alpha_2)$ with a phase shift operator. By doing that, we can overcome the weakness of original QPKE scheme [27].

### III. QEPS-D SIMULATION

The simulation is performed with OptiSystem and the simulation layout is illustrated in Fig. 2. The major modules are explained in the figure caption. The only extra components are needed to discuss here are QEPS and RNG. All others are common for typical coherent optical communications. The random number generator or RNG should be a cryptographic PRNG or pseudo–Quantum Random Number Generator or pQRNG [36] with generated random number meeting cryptographic requirement. pQRNG is capable to take up to 16

KB of the pre-shared secret and produces pseudo random number with excellent randomness [36]. QEPS consists of two operators: $\hat{d}(\alpha_1)$ implemented with standard data modulation such as 16-QAM or QPSK and $\hat{d}(\alpha_2)$ implemented with a random phase shift operator. These two operators together offer a coherent encryption with a generic displacement operator $\hat{d}(\alpha)$. QEPS produces a complex modulation form based on the rand number generated from RNG module. The complex modulation form dictates the signal generator to produce voltages for IQ modulator. In Fig. 2, we omitted the data input which is combined with QEPS. Once the coherent states are generated from CW and pass IQ Modulator, their amplitude and phase would be modulated by IQ modulator then the encrypted cipher coherent states are transmitted over 80 km fiber to coherent detector at the receiver side. Typical coherent detection is applied to produce electrical digital signal and QEPS-d decryption is done before DSP processing. The simulation parameters are given in Table 1.

TABLE 1. SIMULATION PARAMETERS ARE TABULATED.

| | | |
|---|---|---|
| **Layout Parameter** | Sequence length | 65,536 bits |
| | Baud rate | 28 Gbaud |
| | PM period | 1024 |
| **CW Laser and LO Laser** | Center wavelength | 1550 nm |
| | Power | 5 dBm |
| | Linewidth | 0.1 MHz |
| | Azimuth | 0.45 degree |
| **IQ Modulator** | Extinction ratio | 20 dB |
| | Switching bias | 3 V |
| | Insertion loss | 5 dB |
| **EDFA** | Forward pump power | 13-14 mW |
| | Forward pump wavelength | 980 nm |
| | Loss at 1550 nm | 0.1dB/m |
| | Loss at 980 nm | 0.15 dB/m |
| **Optical Fiber** | Length (1 spool) | 80 km |
| | Attenuation | 0.2 dB/km |
| | Dispersion | 0.3 16.75 ps/nm/km |
| | Dispersion slope | 0.4 0.075 ps/nm2/km |
| | Differential group delay | 0.5 0.2ps/km |
| | Effective area | 80 µm² |

We simulated QEPS encryption with the reduced displacement operator for QPSK data modulations and plot constellation diagrams in 3 cases:

1. Constellation right after coherent detection as shown in Fig. 3. This constellation diagram displays the detections of cipher coherent states together with fiber path impacts.

2. Constellation diagram after applying the digital signal processing as shown in Fig. 4.

3. Constellation after applying digital QEPS decryption and DSP compensations as shown in Fig. 5.

Fig. 3 is used to mimic the attacker's coherent detection by assuming the attacker taped good portion of the transmitted cipher coherent signals. Then he/she would obtain a coherent constellation diagram as shown in Fig. 3, which is randomly scattered points. Then we also assume that the attacker knows the data modulation scheme to be QPSK so he/she can apply DSP to compensate and correct the impacts from the fiber path.

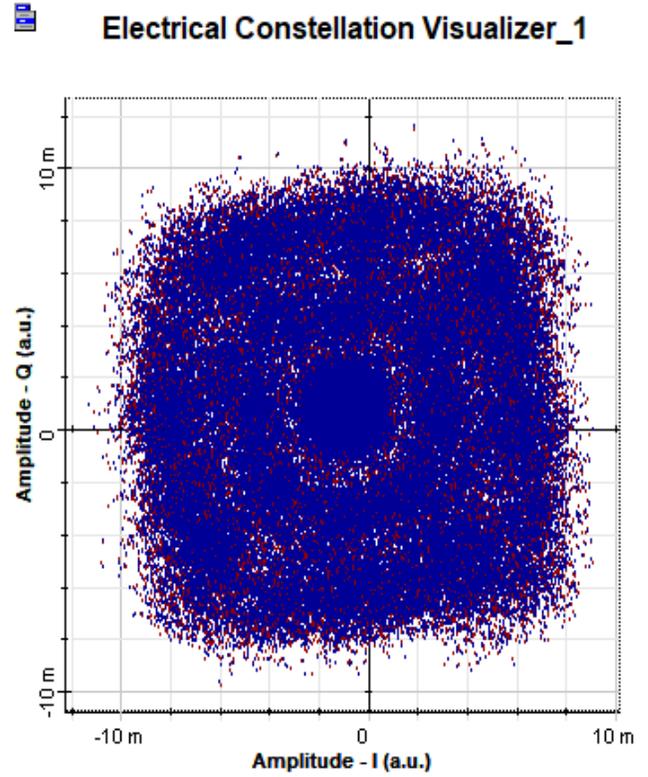

Figure 3. Constellation diagram of directly detected cipher coherent states is displayed.

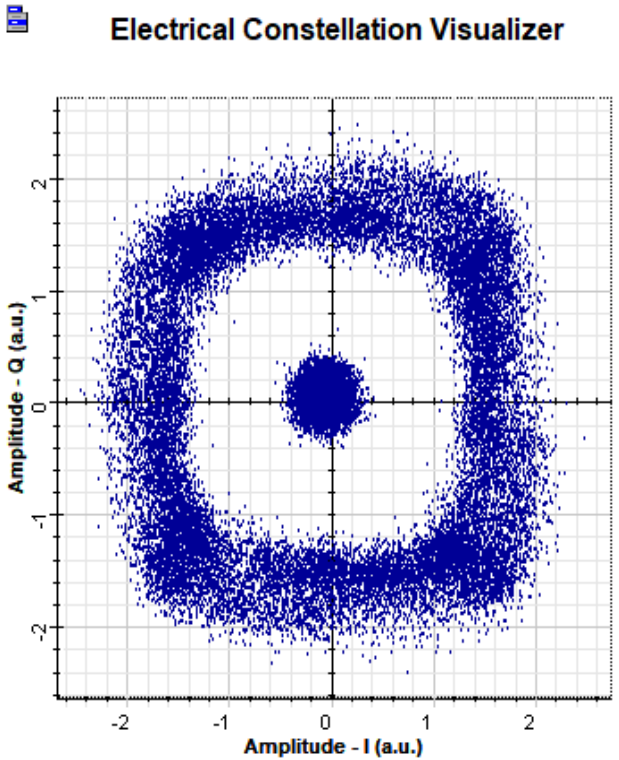

Figure 4. Constellation diagram of directly detected cipher coherent states is displayed after applying the DSP processing. The BER is 0.38.

After applying DSP processing, he/she obtains a constellation diagram as shown in Fig. 4 with a huge Bit-Error-Rate or BER at 0.38. That means, it is impossible to extract any meaningful transmitted data. If we carefully look at Fig. 4, we will notice that there is a square-typed band with 2-unit amplitude, indicating two QPSK modulations through QEPS-d encryption $\hat{d}(\alpha_1)$ on a QPSK data modulation. The square band reflects the phase shift operator $\hat{d}(\alpha_2)$ driving by the random number generated from RNG. The central disk reflects the QPSK data modulations have the opposite phases of $\hat{d}(\alpha_1)$ so they cancel out and give the "zero" amplitudes.

In QPSK data modulation scheme, data values are modulated into phases not in amplitude, so Fig. 4 would not leak transmitted data information. So, they transmission is totally secure.

Coherent detection turns coherent optical domain into coherent electrical domain so digital signal processing can compensate and correct the impacts from the optical path. That is fantastic for QEPS encryption: encryption in coherent optical domain or analogue encryption then decryption in electrical digital domain before DSP processing. That means, QEPS encryption is an analogue encryption which blocks attackers to extract transmitted digital data. Of course, one can apply AES encryption in data then transmit with coherent optical communications which would allow attackers to extract AES ciphertexts. That is the major difference between QEPS and other encryption schemes.

Leveraging the feature of coherent detection, we apply QEPS-d decryption with $\hat{d}(-\alpha)$ driving by the synchronized RNG seeded with the pre-shared secret. Fig. 5 illustrates the constellation diagram with QEPS-d decryption then DSP

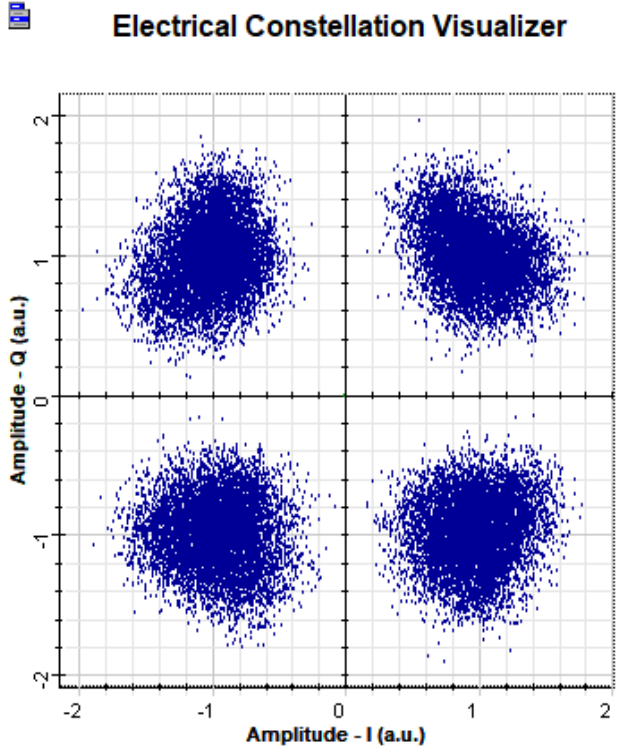

Figure 5. Constellation diagram of QEPS decryption and DSP processing. BER is 0.

processing. It is clearly seen that a QPSK constellation with BER to be zero.

The described technique in the above can be implemented in a round trip as shown in QPKE [27, 30] where Alice becomes Alice Transmission and Alice receiving with a self-shared random secret for encryption and decryption then Bob only performs data modulations, Alice would securely extract Bob's transmitted data without pre-share secret. Using this way, one trick needs to be remembered: phase shift operator must be in a reverse order of transmission side. The round-trip implementation can be also used for true random number distributions, as an alternative of traditional QKD but the key rate can be dramatically increased to 100s gbps. For example, in this simulation, we could achieve 56 gbps with a single polarization and 112 gbps with dual polarizations.

The distance can be extended with EDFA amplification as what we have used in today's coherent optical communications.

IV. CONCLUSION

We briefly introduced QEPS with the reduced displacement operator proposed in [32] and applied it for QPSK data modulation with QPSK implementation of the first displacement operator $\hat{d}(\alpha_1)$ and a randomized phase shift operator of the second displacement operator $\hat{d}(\alpha_2)$. The simulation demonstrates QEPS-d offers security in analogue domain encryption and the transmitted cipher coherent states can not be extracted without knowing the pre-shared secret in symmetric implementation mode. It can be also implemented in a roundtrip scheme without the pre-shared secret which can be used for key distributions over coherent optical communications. The simulation shows that we can achieve 56 gbps distributions rate with a single polarization and 112 gbps with dual polarizations. As what we have demonstrated in [32] that the displacement operator can also be implemented with QAM schemes such as 16-QAM or 32-QAM. That makes QEPS-d be a generic encryption in coherent optical domain or analogue encryption. In the future, we plan to implement it experimentally.